# Dual-Polarization FBMC for Improved Performance in Wireless Communication Systems

Hosseinali Jamal, *Member, IEEE*, David W. Matolak, *Senior Member, IEEE*

*Abstract*— Filter bank multi-carrier (FBMC) offers superior spectral properties compared to cyclic-prefix orthogonal frequency-division multiplexing (CP-OFDM), at the cost of an inherent shortcoming in dispersive channels called intrinsic imaginary interference. In this paper we propose a new FBMC based communication system using two orthogonal polarizations for wireless communication systems: dual-polarization FBMC (DP-FBMC). Using this system we can significantly suppress the FBMC intrinsic interference. Therefore in DP-FBMC all the multicarrier techniques used in CP-OFDM systems such as channel equalization, etc., should be applicable without using the complex processing methods required for conventional FBMC. DP-FBMC also has other interesting advantages over CP-OFDM and FBMC: it is more robust in highly dispersive channels, and also to receiver carrier frequency offset (CFO) and timing offset (TO). In our DP-FBMC system we propose three different structures based on different multiplexing techniques. We show that compared to conventional FBMC, one of these DP-FBMC structures has equivalent complexity and equipment requirements. We compare DP-FBMC with other systems through simulations. According to our results DP-FBMC has potential as a promising candidate for future wireless communication networks.

## I. INTRODUCTION

The orthogonal frequency division multiplexing (OFDM) modulation with the cyclic prefix (CP) extension is at present the most widespread multicarrier communication technique, due to its relative simplicity and robustness against multipath frequency selective channels thanks to the CP. Yet this inserted CP decreases the spectral efficiency, especially in highly-dispersive channels. Also, because of the symbol-time-limited pulses the OFDM spectrum is not compact, and has large spectral sidelobes, and it thus requires a large number of guard subcarriers to reduce the out-of-band power emission, further decreasing spectral efficiency. As an alternative approach to increase the spectral efficiency and offering a more compact power spectral density, filterbank multicarrier (FBMC) has been proposed [1]. The FBMC structure does not require a CP and has very compact spectral shape due to filtering. In many cases this can enhance the spectrum efficiency (throughput) significantly. FBMC has been studied and compared to CP-OFDM for future cellular communication networks such as 5G in [2]-[4]. In the literature several FBMC systems have been proposed and reviewed in recent years. These systems are based on different structures, many of which are listed in [2] and [5]-[7]. In this paper we focus on the most widespread and popular FBMC technique based on Saltzberg's method [8] (known as staggered multitone (SMT) FBMC [5] or OFDM-OQAM). This method makes it possible to have symbol-rate spacing between adjacent subcarriers without intersymbol interference (ISI) and intercarrier interference (ICI) in distortionless channels by introducing a shift of half the symbol period between the in-phase and quadrature components of QAM symbols. Thus in FBMC, the subcarrier symbols are modulated with real offset-QAM (OQAM) symbols and the orthogonality conditions are considered only in the real domain [5]. According to this real orthogonality condition, FBMC incurs a shortcoming due to "intrinsic imaginary interference" in dispersive channels. In the literature there are several proposals for estimating and mitigating intrinsic interference, but all these techniques increase complexity [9]-[18].

Polarization-division multiplexing (PDM) is a physical layer communication technique for multiplexing signals on electromagnetic waves of two orthogonal polarization signal states on the same carrier frequency. This technique has been proposed for microwave links such as satellite television to double the throughput [19], [20]. It has also been proposed for fiber optic communication using two orthogonal left- and right-hand circularly polarized light beams in the same light guide fiber [21], [22]. In terrestrial and air-to-ground (AG) wireless communication environments, due to the non-stability of antenna position and often rich scattering in the wireless channels, using this method (to double throughput) may often not be practical, and would require highly complex receivers to remove the interference resulting from the often small cross-polarization discrimination (XPD). The XPD is a common way of describing the amount by which a channel separates polarizations. It is defined as the ratio of desired polarization mean power to that on the opposite polarization. In this paper, using dual polarization (DP) technique we propose dual-polarization FBMC (DP-FBMC) not to double the capacity but rather to solve the intrinsic imaginary interference shortcoming of FBMC systems in dispersive channels. By using two polarizations in FBMC we basically add another dimension to suppress the intrinsic interference. We show that transmitting symbols on two orthogonal polarizations reduces the interference by a large extent, and in order to further suppress the remaining residual interference we suggest choosing prototype filters with near Nyquist characteristics, such as square-root raised cosine (SRRC) filters.

Using different multiplexing techniques we propose three different DP-FBMC approaches: time-polarization division multiplexing (TPDM), frequency-polarization division multiplexing (FPDM), and time-frequency-polarization division multiplexing (TFPDM). The difference in these methods is the location of transmitted FBMC OQAM symbols in the time, frequency, and polarization domains.

In DP systems, and accordingly in DP-FBMC, the main parameter that should be analyzed is the cross-coupling effect (XPD) of the channel on the received symbols. Electromagnetic wave polarization can change due to various mechanisms (e.g., reflections) as the wave propagates through a channel. Assuming well-designed DP antennas with perfect antenna cross-polarization isolation, the only cross polarization interference arises from channel environment.

Here we briefly provide a literature review for XPD based on both measurement and analytical results provided in [23]-[30]. In [23], [24] the authors describe measurements and analysis for the 1800 MHz frequency band, whereas in [25]-[27] there are some results for 2.5 GHz, and in [28], [29] results for indoor mmWave bands at 28 and 73 GHz. According to the 1800 MHz results, the XPD ranges from 5 to 15 dB. The largest XPD values pertain to outdoor LOS-like channels and the smallest occurs in more rich scattering NLOS channels (both indoor and outdoor). Results for the 2.5 GHz band show a range of XPD values similar to the 1800 MHz band. In the mmWave bands the XPD values are significantly larger than those at the lower frequencies. According to the results at 28 GHz, the XPD values are in the range of 8-14 dB and 14-24 dB for NLOS and LOS cases, and at 73 GHz, XPD ranges are 13-18 dB and 21-31 dB for NLOS and LOS cases, respectively. The authors in [30] did an extensive literature overview of experimental data regarding DP channels. According to their review of empirical data from different references, the XPD results from channel effects was measured between 4 to 8 dB in NLOS outdoor cases, up to 15 to 19 dB is LOS urban and rural areas, 3 to 8 dB in NLOS indoor cases, and up to 15 dB in LOS indoor scenarios. In this paper our main contribution is to remove the intrinsic interference in FBMC systems by multiplexing symbols on DP antennas, and in our DP analysis and structures we do not use polarization diversity or spatial multiplexing.

The remainder of this paper is organized as follows: in Section II we describe the OFDM-OQAM based FBMC system model. In Section III we describe our proposed DP-FBMC communication systems, and through analysis we describe the cross-coupling effect on the BER performance. In Section IV we provide the simulation results and compare CP-OFDM, conventional FBMC and DP-FBMC systems' performance in four different communication channel scenarios: an air-to-ground (AG) channel based on NASA measurements, and the pedestrian "channels A, B" and vehicular "channel B" from ITU recommendations. We also compare power spectral density (PSD), and evaluate the performance degradation in low XPD conditions. In Section V we provide conclusions and suggested future work.

## II. FBMC SYSTEM MODEL

In the OFDM-OQAM form of FBMC, real valued OQAM symbols $a_{n,m}$ are filtered through prototype filter $h(t)$ and then modulated across *N* subcarriers as described by the following continuous form equation,

$$x(t) = \sum_{n=0}^{N-1} \sum_{m \in \mathbb{Z}} a_{n,m} h\left(t - m\frac{T}{2}\right) e^{\frac{j2\pi nt}{T}} e^{j\theta_{n,m}}. \quad (1)$$

Prototype filter $h(t)$ is a finite impulse response filter with a length *L=KN*, with *K* defined as the overlapping factor. In this equation $\theta_{n,m} = \frac{\pi}{2}(n+m)$ is a phase term between adjacent subcarriers and symbols to satisfy the orthogonality condition in the real domain at the receiver [5], [6]. According to (1) symbols are offset or overlapped by half a symbol duration, *T*/2. For implementation, to reduce the complexity, a polyphase network (PPN) of prototype filters and fast and inverse fast Fourier transforms (FFT, IFFT) are used, as shown in Figure 1. For more details regarding the PPN structure and FFT implementation refer to [2], [5], or [31]. In Figure 1(a), for the FBMC transmitter, note that after the π/2 phase shifts, the IFFT input symbols are either purely real or purely imaginary values. After the IFFT block, subcarriers will be filtered through the PPN network, and for each block of *N* input subcarriers, what comes out of the parallel to serial (P/S) conversion is a signal vector with the same length as the prototype filter. These symbol vectors are then overlapped or offset by *N*/2 to achieve maximum spectral efficiency.

In Figure 2 we depict a useful diagram called time-frequency phase-space lattice to illustrate the transmitted symbols in time, frequency, and phase. This figure shows the time-frequency lattice of FBMC symbols for an example of 16 subcarriers. Note that all symbols adjacent in time or frequency have a π/2 phase shift between them (adjacent solid circles and squares) to satisfy the real orthogonality condition [5], thus in perfect (distortionless) channel conditions there is no ISI or ICI at the receiver. As mentioned, one main shortcoming of FBMC compared to OFDM emanates from this real orthogonality, which will be violated in non-perfect channel conditions.

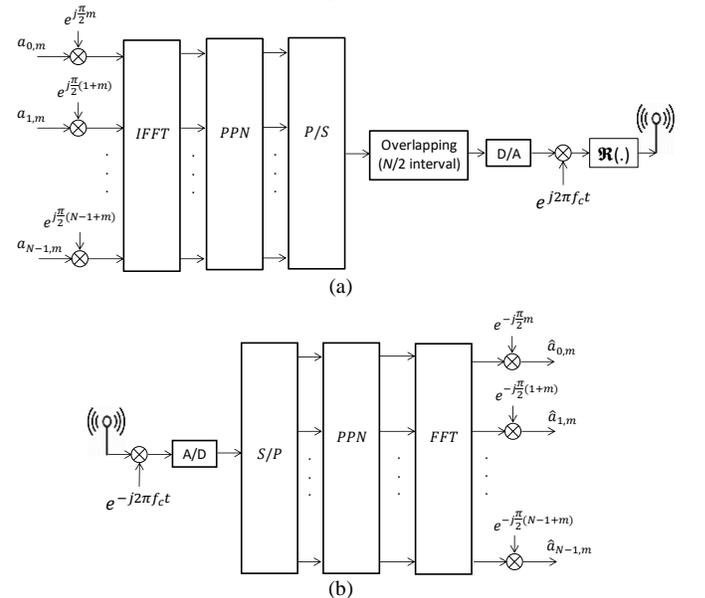

Figure 1. OQAM-OFDM (FBMC) communication system; (a) transmitter, (b) receiver.

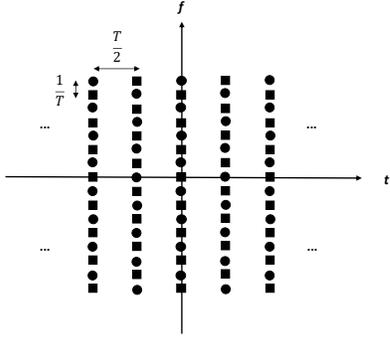

Figure 2. FBMC symbols time-frequency phase-space lattice ($N=16$). Circles and squares denote a relative $\pi/2$ phase shift between symbols adjacent in time and/or frequency.

This problem yields what is called intrinsic imaginary interference, and this makes the use of the straightforward OFDM channel equalization and MIMO techniques impractical in FBMC. In order to reduce this interference for channel equalization and MIMO purposes, several methods have been proposed in recent years. Among these techniques are scattered or auxiliary pilots [10], [11], preamble-based channel estimation [12], spreading techniques for MIMO applications [13], and per-subchannel equalizers based on the frequency sampling approach for multi-antenna receivers [14].

Some of these methods add extra computational complexity at receivers and require data payloads. In this paper we show that in DP-FBMC systems we can suppress the intrinsic imaginary interference very effectively without any extra processing and data payload.

Here first we analyze the intrinsic imaginary interference in conventional FBMC since this is useful to explain DP-FBMC as well. First we re-write (1) in the discrete form as follows,

$$x[k] = \sum_{n=0}^{N-1} \sum_{m \in \mathbb{Z}} a_{n,m} h\left[k - m\frac{N}{2}\right] e^{\frac{j2\pi n(k-\frac{L}{2})}{N}} e^{j\theta_{n,m}}. \quad (2)$$

Now we can rearrange (2) as follows,

$$x[k] = \sum_{n=0}^{N-1} \sum_{m \in \mathbb{Z}} a_{n,m} Q_{n,m}[k], \quad (3)$$

where,

$$Q_{n,m}[k] = h\left[k - m\frac{N}{2}\right] e^{\frac{j2\pi n(k-\frac{L}{2})}{N}} e^{j\theta_{n,m}}. \quad (4)$$

Here the $Q_{n,m}[k]$ functions are the time- and frequency-shifted versions of the prototype filter. In the case of an ideal channel (only considering the transceiver response), the demodulated symbol over the $n'$th subcarrier and the $m'$th instant is determined using the inner product of $x[k]$ and $Q_{n',m'}[k]$ as follows,

$$a_{n',m'} = \langle x, Q_{n',m'} \rangle = \sum_{k=-\infty}^{+\infty} x[k] Q^*_{n',m'}[k]$$
$$= \sum_{k=-\infty}^{+\infty} \sum_{n=0}^{N-1} \sum_{m \in \mathbb{Z}} a_{n,m} Q_{n,m}[k] Q^*_{n',m'}[k]. \quad (5)$$

In order to perfectly detect the transmitted symbols without any errors (such that $a_{n',m'}=a_{n,m}$), we require one orthogonality condition. Hence assuming a perfect distortionless channel, and with $\theta_{n,m}$ as described in (1), we require the real orthogonality condition as follows,

$$\Re\{\langle Q_{n,m}, Q_{n',m'} \rangle\} = \Re\left\{\sum_{k \in \mathbb{Z}} Q_{n,m}[k] Q^*_{n',m'}[k]\right\} = \delta_{n,n'}\delta_{m,m'}, \quad (6)$$

where $\delta_{n,n'}$ is the Kronecker delta, equal to 1 if $n=n'$ and 0 if $n \neq n'$. Now considering the channel and AWGN (we note that noise is not strictly white, but for all practical filters is very nearly white), the received symbols can be written as follows,

$$r_{n',m'} = h_{n',m'} a_{n',m'} + \gamma_{n',m'}$$
$$+ \underbrace{\sum_{(n,m) \neq (n',m')} h_{n,m} a_{n,m} \sum_{k=-\infty}^{+\infty} Q_{n,m}[k] Q^*_{n',m'}[k]}_{I_{n',m'}}, \quad (7)$$

where $h_{n',m'}$ is the complex channel coefficient at subcarrier $n'$ and time index $m'$, and the term $I_{n',m'}$ is the intrinsic interference. The $\gamma_{n',m'}$ term is the noise variable. In practice, having well-localized filters, most of the energy of the filter impulse response is localized in a restricted region around the considered symbol $(n', m')$ [4], [11]. Consequently, we assume the considered intrinsic interference is confined only to this restricted set (denoted as $\vartheta_{n',m'}$). Also, assuming the channel is constant over this summation zone, which is often valid for a variety of practical channels [11], we can write,

$$r_{n',m'} \approx h_{n',m'}(a_{n',m'} + \hat{I}_{n',m'}) + \gamma_{n',m'}, \quad (8)$$

where $\hat{I}_{n',m'}$ is the intrinsic interference due to the restricted set of symbols and is calculated as follows,

$$\hat{I}_{n',m'} = \sum_{(n,m) \in \vartheta_{n',m'}} a_{n,m} \sum_{k=-\infty}^{+\infty} Q_{n,m}[k] Q^*_{n',m'}[k], \quad (9)$$

and,

$$\vartheta_{n',m'} = \{\forall (n,m) | n \neq n', m \neq m', |n - n'| \leq \Delta n, |m - m'| \leq \Delta m, h_{n',m'} \cong h_{n,m}\}. \quad (10)$$

The term $\vartheta_{n',m'}$ is the set of nearby indices $(n,m)$ within $\Delta n$ subcarriers and $\Delta m$ symbols of the reference subcarrier and symbol indices $(n', m')$ where the channel has constant response $h_{n',m'} \cong h_{n,m}$. For many practical well-localized prototype filters, $\Delta n, \Delta m$ can be as small as one [11]. According to (6) and because the transmitted OQAM symbols are real-valued, the intrinsic interference $\hat{I}_{n',m'}$ is purely imaginary, and this is why it is called imaginary intrinsic interference.

As long as $\hat{I}_{n',m'}$ (which can be seen as a 2D-ISI) is unknown at the receiver the application of pilot scattering channel estimation and therefore MIMO are extremely complex. Therefore for channel equalization and MIMO applications we must mitigate this interference. In [10] and later in [11] the authors proposed the use of auxiliary pilot symbols at the transmitter adjacent to actual channel estimation pilots: these auxiliary symbols are allocated to effectively remove $\hat{I}_{n',m'}$ interference. For calculating the $\hat{I}_{n',m'}$ values we define the filter time-frequency localization function as follows,

$$Q_{n,m}^{n',m'} = \langle Q_{n,m}, Q_{n',m'} \rangle = \sum_{k=-\infty}^{+\infty} Q_{n,m}[k] Q_{n',m'}^*[k]. \quad (11)$$

By these calculations and knowing the purely real or imaginary symbols surrounding the transmitted symbols on $\vartheta_{n',m'}$, we can calculate the intrinsic interference from (9).

## III. PROPOSED DP-FBMC SYSTEM MODEL

In Figure 3 we illustrate the dual polarization communication system using vertical and horizontal polarization antennas. In our DP-FBMC proposal we describe three different multiplexing approaches. In Figure 4 we depict the time-frequency-polarization phase-lattice of all DP-FBMC structures, where blue and red colored symbols representing transmitting symbols on vertical and horizontal polarizations, respectively. In Figure 4(a) DP-FBMC *Structure I* based on TPDM is depicted. In this method we separate or isolate adjacent symbols on two orthogonal polarizations by multiplexing symbols in time. By this approach we can remove the intrinsic interference that results from (temporally) adjacent symbols. Interference still exists from symbols on nearby subcarriers with much lower power. Here we note (but do not provide the results for brevity) that this structure could also be used on CP-OFDM with a similar BER performance advantage as DP-FBMC in highly dispersive channels (as will be shown later). Yet because OFDM has pulse length equal to the symbol spacing, DP multiplexing will result in temporal gaps between transmitting symbols on each polarization, which significantly degrades the peak-to-average power ratio (PAPR).

In the DP-FBMC *Structure II* based on FPDM, as shown in Figure 4(b), we separate or isolate the adjacent transmitting subcarriers on two polarizations by multiplexing symbols in frequency. This method is not as useful as the first and third structures in removing intrinsic interference because as we will explain shortly, most of the intrinsic interference comes from directly adjacent symbols on the same subcarrier index (at the same frequency, i.e., adjacent symbols on same row). In Figure 4(c) we depict the time-frequency-polarization phase-lattice structure of DP-FBMC *Structure III* based on TFPDM. In this structure we transmit two halves of the OQAM symbols on two orthogonal polarizations at every symbol time, and then subsequently switch the order of half the subcarriers on the two polarizations at the next symbol time. Hence if polarization isolation is perfect, the majority of the intrinsic imaginary interference (from nearest neighbor symbols) will be removed. Here we briefly mention that there are also some structures using complex QAM symbols (this can be done by dividing transmitted QAM symbols on even and odd subcarriers of each polarization), and according to our simulations (not shown in this paper) we determine that OQAM modulation has more robustness to the polarization cross-coupling thanks to the $\pi/2$ phase difference between adjacent symbols by applying $\theta_{n,m}$ in (1).

Therefore in this paper we only analyze the DP-FBMC systems based on OQAM modulation. Here we also note that *Structures II* and *III* could be used in CP-OFDM, but for brevity we will not show the results of DP CP-OFDM; we simply note that DP CP-OFDM has similar BER results as DP-FBMC, but without the other DP-FBMC advantages such as better spectral efficiency.

In order to provide the numerical analysis and comparison of prototype filters, using (11) and considering the restricted set $(\vartheta_{n',m'})$, in Tables 1-5 we list the $Q_{n,m}^{n',m'}$ values for $\Delta n=2$, $\Delta m=3$ for the isotropic orthogonal transform algorithm (IOTA) [1], PHYDYAS [33], and squared-root raised cosine (SRRC) prototype filters. As explained the most power of intrinsic interference comes from immediate neighboring symbols, especially adjacent symbols in the same subcarrier index (dark shaded cells in tables).

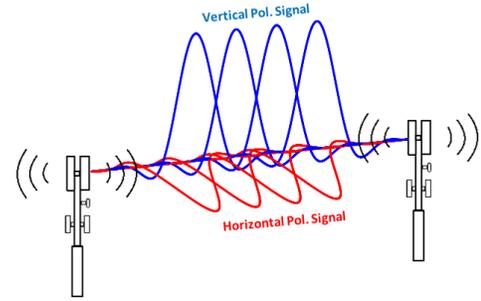

Figure 3. DP-FBMC wireless communication link (*Structure I*).

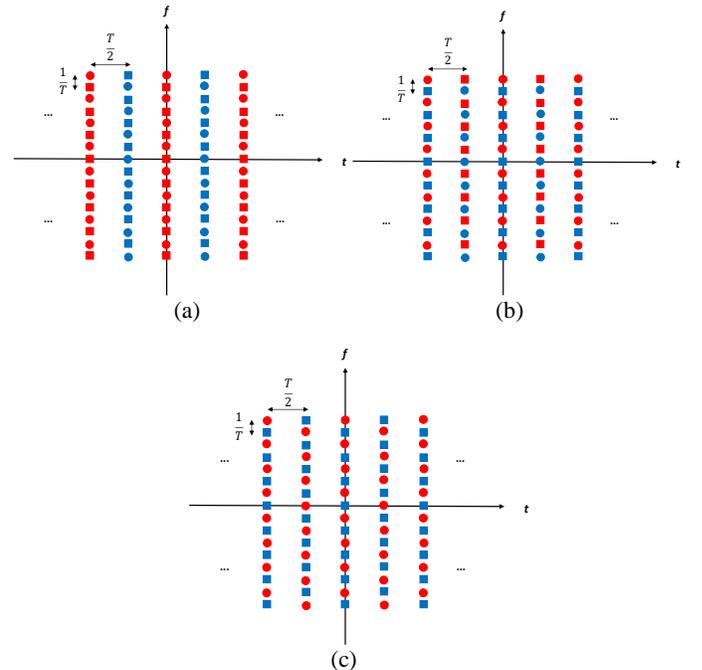

Figure 4. DP-FBMC symbols time-frequency-polarization phase-lattice, (a) *Structure I* based on TPDM, (b) *Structure II* based on FPDM, (c) *Structure III* based on TFPDM. Blue and red colored symbols represent symbols on V and H polarizations, respectively. Solid squares and circles represent $\pi/2$ phase difference between adjacent symbols (real orthogonality condition).

Table 1. $Q_{n,m}^{n',m'}$ values using IOTA filter $K$=4.

| $(n,m)$ | $m=m'$-3 | $m=m'$-2 | $m=m'$-1 | $m=m'$ | $m=m'$+1 | $m=m'$+2 | $m=m'$+3 |
|---|---|---|---|---|---|---|---|
| $n=n'$-2 | 0.0194j | 0 | -0.0413j | 0 | 0.0413j | 0 | 0.0194j |
| $n=n'$-1 | -0.0116j | -0.0413j | -0.2327j | -0.4378j | -0.2327j | -0.0413j | -0.0116j |
| $n=n'$ | 0.0194j | 0 | -0.4380j | 1 | 0.4380j | 0 | 0.0194j |
| $n=n'$+1 | -0.0116j | 0.0413j | -0.2327j | 0.4378j | -0.2327j | 0.0413j | -0.0116j |
| $n=n'$+2 | 0 | 0 | -0.0413j | 0 | 0.0413j | 0 | 0 |

Table 2. $Q_{n,m}^{n',m'}$ values using PHYDYAS filter $K$=4.

| $(n,m)$ | $m=m'$-3 | $m=m'$-2 | $m=m'$-1 | $m=m'$ | $m=m'$+1 | $m=m'$+2 | $m=m'$+3 |
|---|---|---|---|---|---|---|---|
| $n=n'$-2 | 0.064j | 0 | 0 | 0 | 0 | 0 | 0.064j |
| $n=n'$-1 | -0.044j | -0.125j | -0.205j | -0.239j | -0.205j | -0.125j | -0.044j |
| $n=n'$ | 0.064j | 0 | -0.564j | 1 | 0.564j | 0 | 0.064j |
| $n=n'$+1 | -0.044j | 0.125j | -0.205j | 0.239j | -0.205j | 0.125j | -0.044j |
| $n=n'$+2 | 0 | 0 | 0 | 0 | 0 | 0 | 0 |

Table 3. $Q_{n,m}^{n',m'}$ values using SRRC filter $K$=4.

| $(n,m)$ | $m=m'$-3 | $m=m'$-2 | $m=m'$-1 | $m=m'$ | $m=m'$+1 | $m=m'$+2 | $m=m'$+3 |
|---|---|---|---|---|---|---|---|
| $n=n'$-2 | 0.1122j | 0 | 0 | 0 | 0 | 0 | 0.1122j |
| $n=n'$-1 | -0.095j | -0.1263j | -0.15j | -0.1589j | -0.15j | -0.1260j | -0.095j |
| $n=n'$ | 0.1122j | 0 | -0.6015j | 1 | 0.6015j | 0 | 0.1122j |
| $n=n'$+1 | -0.095j | 0.1263j | -0.15j | 0.1589j | -0.15j | 0.1260j | -0.095j |
| $n=n'$+2 | 0 | 0 | 0 | 0 | 0 | 0 | 0 |

Table 4. $Q_{n,m}^{n',m'}$ values using SRRC filter $K$=8.

| $(n,m)$ | $m=m'$-3 | $m=m'$-2 | $m=m'$-1 | $m=m'$ | $m=m'$+1 | $m=m'$+2 | $m=m'$+3 |
|---|---|---|---|---|---|---|---|
| $n=n'$-2 | 0.1857j | 0 | 0 | 0 | 0 | 0 | 0.1857j |
| $n=n'$-1 | -0.0646j | -0.0695j | -0.0725j | -0.0735j | -0.072j | -0.0694j | -0.0646j |
| $n=n'$ | 0.1857j | 0 | -0.6278j | 1 | 0.627j | 0 | 0.1857j |
| $n=n'$+1 | -0.0646j | 0.0695j | -0.0725j | 0.0735j | -0.072j | 0.0694j | -0.0646j |
| $n=n'$+2 | 0 | 0 | 0 | 0 | 0 | 0 | 0 |

For DP-FBMC we turned to the classic SRRC filter. Via some numerical trials, we determined heuristically that a roll-off factor $\alpha$=2/$K$ can optimize the intrinsic interference reduction. As shown in Tables 3 and 4 we can see that using the suggested SRRC filter and larger overlapping factors we can decrease the filter localization (hence intrinsic interference) on surrounding symbols. Therefore choosing the suggested SRRC filter, especially with larger overlapping factors such as $K$=8 or higher, significantly reduces the filter response samples (hence intrinsic interference).

Also from these tables and as mentioned before we recognize that the majority of the intrinsic interference results from the temporally adjacent symbols on the same subcarrier ($n = n'$ and $m = m'$-1, or $m = m'$+1) and this is exactly the reason why the DP-FBMC *Structure II* is not effective in removing the intrinsic interference. Hence if *Structure II* is used, even with dual polarization we need intrinsic interference cancelation techniques for channel equalization such as those in conventional FBMC. Henceforth we only show results for *Structures I* and *III*.

Following, in (12)-(14) we can write the multiplexed OQAM symbols for DP-FBMC *Structures I*, *II*, and *III*, respectively,

$$a_{n,m}^H = \begin{cases} a_{n,m} & m\ even \\ 0 & m\ odd \end{cases}$$
$$a_{n,m}^V = \begin{cases} a_{n,m} & m\ odd \\ 0 & m\ even \end{cases} \quad (12)$$

$$a_{n,m}^H = \begin{cases} a_{n,m} & n\ even \\ 0 & n\ odd \end{cases}$$
$$a_{n,m}^V = \begin{cases} a_{n,m} & n\ odd \\ 0 & n\ even \end{cases} \quad (13)$$

$$a_{n,m}^H = \begin{cases} a_{n,m} & m\ even, n\ even \\ 0 & m\ even, n\ odd \\ 0 & m\ odd, n\ even \\ a_{n,m} & m\ odd, n\ odd \end{cases}$$

$$a_{n,m}^V = \begin{cases} 0 & m\ even, n\ even \\ a_{n,m} & m\ even, n\ odd \\ a_{n,m} & m\ odd, n\ even \\ 0 & m\ odd, n\ odd \end{cases} \quad (14)$$

Now using these expressions we can write the transmitted waveforms on each polarization according to (15). Note that we can also use circular right-handed and left-handed (or any other) orthogonal polarizations, but here we use the *H* and *V* notations for horizontal and vertical polarizations.

$$x^H[k] = \sum_{n=0}^{N-1} \sum_{m \in \mathbb{Z}} a_{n,m}^H h\left[k - m\frac{N}{2}\right] e^{\frac{j2\pi n\left(k-\frac{L}{2}\right)}{N}} e^{j\theta_{n,m}}$$

$$x^V[k] = \sum_{n=0}^{N-1} \sum_{m \in \mathbb{Z}} a_{n,m}^V h\left[k - m\frac{N}{2}\right] e^{\frac{j2\pi n\left(k-\frac{L}{2}\right)}{N}} e^{j\theta_{n,m}} \quad (15)$$

Here we briefly compare the complexity of these structures with that of conventional FBMC. First considering the direct equation forms of (2) and (15), we find that in DP-FBMC *Structures II* and *III*, for each symbol period, the number of multiplications is reduced by a factor of two on each polarization as long as the input symbols on half the subcarriers are zero. Therefore the complexity of the DP-FBMC transmitter is similar to that of conventional FBMC. DP-FBMC *Structure I* also has complexity similar to that of conventional FBMC (based on the direct form).

If though we look at the fast implementation of the systems based on IFFT, FFT, and PPN implementation, for DP-FBMC *Structures II* and *III*, first we deduce that we need a second IFFT and FFT as well as second PPN at both transmitter and receiver, second we note that at every symbol time half of the subcarrier samples are zero so only half the subcarrier samples are needed, therefore we can use the pruned IFFT/FFT algorithms [34]-[36] to reduce the added complexity. Based on Skinner's algorithms [35], pruning the vector of input samples with length $N$/2 for an $N$-point IFFT requires $2Nlog_2(N/2)$ real multiplications and $3Nlog_2(N/2) + N$ real additions. Based on Markel's algorithm [34] pruning output samples with length $N$/2 of an $N$-point FFT requires $2Nlog_2(N/4)$ real multiplications and $3Nlog_2(N/2)$ real additions [9]. The pruned IFFT/FFT is effective for a small number of subcarriers (e.g., less than 32), but for a large number of subcarriers this complexity reduction is not effective. Also after IFFT/FFT processing (*Structures II* and *III*), for PPN filtering we need twice the multiplications of conventional FBMC. Therefore DP-FBMC *Structures II* and *III* have higher complexity than conventional FBMC.

For *Structure I*, as long as we can share the same IFFT/FFT at every symbol period and polarization, we have the same complexity as conventional FBMC, therefore we suggest and further study *Structure I* as our main DP-FBMC structure. A complete complexity analysis is reserved for future work.

Regarding the transmit power in all structures, as long as half the symbols are nulled accordingly, each DP-FBMC antenna employs half the power of conventional FBMC, hence lower cost power amplifiers may be used. Received SNR or the energy per bit ($E_b$) to noise density ratio $E_b/N_0$ remains constant.

For cross-coupling analysis on DP-FBMC, first we define the XPD according to [44] as follows,

$$XPD = \frac{E\{|h^{VV}|^2\}}{E\{|h^{HV}|^2\}} = \frac{E\{|h^{HH}|^2\}}{E\{|h^{VH}|^2\}}, \quad (16)$$

where a symmetric leakage is assumed. This "symmetry" assumption was made for V/H polarizations [35] and also concluded by the measurements reported in [36] where the leakage from polarization V to H and H to V have the same average power. In (16) $h^{VV}$ and $h^{HH}$ are the narrowband co-polarization channel responses between (co-) polarized antennas and $h^{HV}$, $h^{VH}$ are the cross-polarized channel responses.

Assuming the symmetric channel model and symmetric structure of DP-FMBC we can further extend (7) for H polarization as follows (similarly for polarization V),

$$r_{n',m'}^H = h_{n',m'}^{HH} a_{n',m'}^H + \gamma_{n',m'}^H$$
$$+ \underbrace{\sum_{(n,m)\in\vartheta_{n',m'}^H} h_{n,m}^{HH} a_{n,m}^H \sum_{k=-\infty}^{+\infty} Q_{n,m}[k] Q_{n',m'}^*[k]}_{I_{n',m'}^H}$$
$$+ \underbrace{\sum_{(n,m)\in\vartheta_{n',m'}^V} h_{n,m}^{VH} a_{n,m}^V \sum_{k=-\infty}^{+\infty} Q_{n,m}[k] Q_{n',m'}^*[k]}_{I_{n',m'}^V}, \quad (17)$$

where $h_{n,m}^{HH}$ is the complex co-polarization channel coefficient at subcarrier $n$ and time index $m$, and $h_{n,m}^{VH}$ is the complex cross-polarization channel coefficient at subcarrier $n$ and time index $m$. Thus the term $I_{n',m'}^H$ is defined as an intrinsic interference caused from the co-polarization symbols, and $I_{n',m'}^V$ is the intrinsic interference caused from the cross-polarization symbols. Note that in the case of perfect XPD, the intrinsic interference from the cross-polarization antenna can be mitigated, hence the only remaining interference is caused by the transmitted symbols on co-polarized antennas.

Similar to the uni-polarization FBMC analysis (assuming well-localized filters) most of the energy of the filter impulse response is localized in a restricted set around the considered symbol. Consequently, we assume the considered intrinsic interference is confined only on the restricted set (denoted as $\vartheta_{n',m'}^H$, $\vartheta_{n',m'}^V$ for H and V polarizations, respectively). Therefore, assuming the channel is constant for this summation zone, for H polarization symbols (and similarly for V polarization) we can write,

$$r_{n',m'}^H \approx h_{n',m'}^{HH}(a_{n',m'}^H + \hat{I}_{n',m'}^H) + h_{n',m'}^{VH} \hat{I}_{n',m'}^V + \gamma_{n',m'}^H \quad (18)$$

where,

$$\hat{I}_{n',m'}^H = \sum_{(n,m)\in\vartheta_{n',m'}^H} a_{n,m}^H \sum_{k=-\infty}^{+\infty} Q_{n,m}[k] Q_{n',m'}^*[k],$$

$$\hat{I}_{n',m'}^V = \sum_{(n,m)\in\vartheta_{n',m'}^V} a_{n,m}^V \sum_{k=-\infty}^{+\infty} Q_{n,m}[k] Q_{n',m'}^*[k]. \quad (19)$$

Depending on the DP-FBMC structure that we chose, $\vartheta_{n',m'}^H$ and $\vartheta_{n',m'}^H$ sets can be defined, accordingly. For example, in DP-FBMC *Structure I* and assuming using a well-localized filter, and $\Delta n = 1$ and $\Delta m = 1$ we have,

$$\vartheta_{n',m'}^H = \{\forall (n,m) | m = m', n \neq n', |n - n'| \leq 1, h_{n',m'}^{HH} \cong h_{n,m}^{HH}\},$$
$$\vartheta_{n',m'}^V = \{\forall (n,m) | m \neq m', |m - m'| \leq 1, |n - n'| \leq 1, h_{n',m'}^{VH}$$
$$\cong h_{n,m}^{VH}\} \quad (20)$$

Therefore comparing (20) with uni-polarized FBMC (8), assuming non-ideal XPD case, the total intrinsic interference ($\hat{I}_{n',m'}$ in uni-polarized FBMC) is divided on two polarization domains where the term $h_{n',m'}^{VH} \hat{I}_{n',m'}^V$ is the interference from the V antenna polarization to the H polarization. In a perfect XPD situation the only interfering part is the co-polarized intrinsic interference $h_{n',m'}^{HH} \hat{I}_{n',m'}^H$, and according to (19) and depending on filter type (for example suggested SRRC with larger overlapping factors) this interference can be significantly reduced. For cross-coupling scenarios, using practical XPD values, and as will be seen in the BER results, the intrinsic interference due to the non-ideality of XPD is tolerable (even in highly frequency selective channels with XPDs as small as 3 dB), and DP-FBCM yields performance similar to conventional FBMC and CP-OFDM.

## IV. SIMULATION RESULTS

In this section we compare the performance of CP-OFDM, conventional FBMC, and DP-FBMC via MATLAB simulations. We evaluate BER performance in different example channels, and the effects of carrier time and frequency offsets. We also compare the PSD of DP-FBMC using different prototype filters and overlapping factors. In addition, we evaluate the performance of DP-FBMC in the presence of polarization angular mismatch, as well as BER vs. XPD. At the end we compare the PAPR results.

In Figure 5 plots, we show the BER vs. $E_b/N_0$ for these communication systems with 16-QAM modulation order, for four example channels. Here we note that we use *Structures I* (which has identical result as *Structure III*). In these simulations there is no channel coding and we chose $N$=512 subcarriers, 16 symbols per frame, and a channel bandwidth $B$=10 MHz.

For the multipath channel fading models we have four different tapped delay line models for four different environment scenarios. The first channel model is a simple over-water strong line of sight (LOS) air-to-ground (AG) channel model based on NASA measurement results [37]. The next three channels are the pedestrian A, B and vehicular channel B from ITU-R Recommendation M.1225 [38]. In Table 5 we list the multipath power delay profiles for these channel models along with root-mean-square delay-spread (RMS-DS) values and fading models. In our analysis and BER performance simulation results, these channels represent mildly-dispersive for AG and pedestrian A, dispersive for pedestrian B, and highly-dispersive for vehicular B channels.

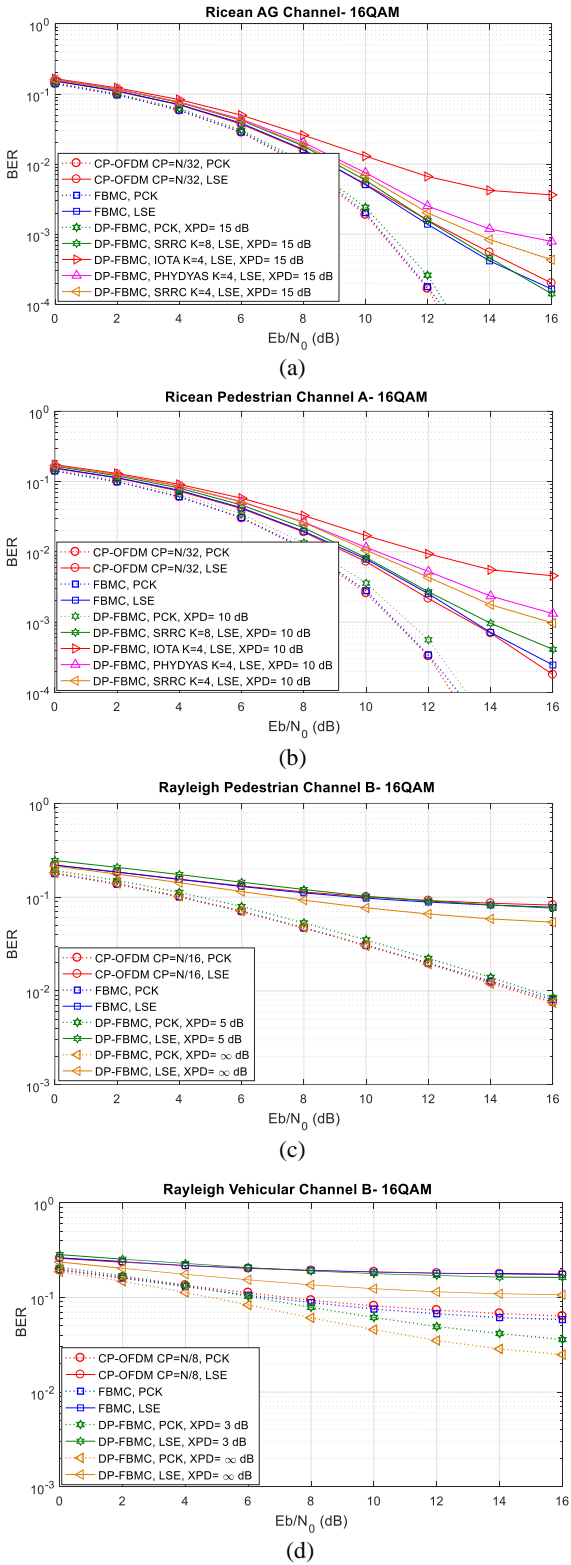

Figure 5. BER vs. $E_b/N_0$ with least-square equalization (LSE) and perfect channel knowledge (PCK) channel equalization, 16-QAM; (a) AG channel (b) ITU *pedestrian* A channel (c) *ITU pedestrian* B channel, and (d) *ITU vehicular* B channel

As typical minimum XPD values in these channels, from [30] we chose 15, 10, 5, and 3 dB for AG, pedestrian A, pedestrian B, and vehicular B channels, respectively.

Table 5. Power delay profile, RMS-DS values, and fading models of example channel models.

| Tap | AG LOS Channel | | Pedestrian Channel A | | Pedestrian Channel B | | Vehicular Channel B | |
|---|---|---|---|---|---|---|---|---|
| | $\tau$ (ns) | $\bar{P}$ (dB) | $\tau$ (ns) | $\bar{P}$ (dB) | $\tau$ (ns) | $\bar{P}$ (dB) | $\tau$ (ns) | $\bar{P}$ (dB) |
| 1 | 0 | 0 | 0 | 0 | 0 | 0 | 0 | -2.5 |
| 2 | 45 | -12 | 110 | -9.7 | 200 | -0.9 | 300 | 0 |
| 3 | 200 | -22.3 | 190 | -19.2 | 800 | -4.9 | 8900 | -12.8 |
| 4 | | | 410 | -22.8 | 1200 | -8 | 12900 | -10 |
| 5 | | | | | 2300 | -7.8 | 17100 | -25.2 |
| 6 | | | | | 3700 | -23.9 | 20000 | -16 |
| RMS-DS (ns) | $\cong 18$ | | $\cong 46$ | | $\cong 633$ | | $\cong 4000$ | |
| Fading | Ricean (Rice factor 30 dB) | | Ricean (Rice factor 10 dB) | | Rayleigh | | Rayleigh | |

We use Ricean fading with Rice factor 30 dB for the strong LOS AG channel. For the pedestrian A channel, the first tap has Ricean fading with Rice factor 10 dB, with the remaining taps incurring Rayleigh fading. All taps in the pedestrian B and vehicular A channel incur Rayleigh fading. In our simulations the transmitted signal is subject to slow fading for all cases. For example, at a 5 GHz carrier frequency and maximum velocity of 300 m/s for the AG case, the maximum Doppler shift is $f_D = v/\lambda = 5$ kHz. Doppler spreads for the slower moving terrestrial platforms are orders of magnitude smaller. The channel coherence time, denoted $T_c$, is inversely proportional to Doppler spread, therefore for the AG case, $T_c \cong 0.2$ ms. Thus as long as our 10 MHz bandwidth signal sample period is much smaller than $T_c$, the transmitted symbols are subjected to slow fading. In BER simulations we assume that any Doppler shifts are tracked and fully compensated at the receiver.

In the CP-OFDM transmitter, we ensure that the CP length is longer than the maximum delay spread of the multipath fading channel: this yields 1/32 of symbol period for the AG and pedestrian channel A, and 1/16 of symbol period for pedestrian B and 1/8 for vehicular channel A. In all communication systems, we use 33 subcarriers as a typical number for guard band (17 on the left and 16 on the right of the signal spectrum), and also use a null DC subcarrier at the center of the spectrum. In these BER results we also show the results using perfect channel knowledge and zero-forcing equalization for comparison.

For channel estimation we used 30 equally spaced subcarriers every 4 symbol periods as scattered pilots in all systems. For this pilot-based channel estimation, we used least square (LS) and discrete Fourier transform (DFT)-based interpolation techniques [39]. For the pilot–based channel estimation in conventional FBMC we used the auxiliary pilot technique based on [10], and assigned 1 auxiliary pilot symbol adjacent to each pilot symbol, and we chose $\Delta n = 2$ and $\Delta m = 2$ for calculating and removing the intrinsic interference. Note that the total number of pilot symbols (including auxiliary symbols in FBMC) for channel equalization in all systems is the same, hence number of data symbols of all systems are identical. For DP-FBMC the auxiliary pilot symbols of conventional FBMC are allocated on the other polarization for channel equalization purpose, thus FBMC and DP-FBMC have the same number of allocated symbols for channel equalization.

According to the BER results, DP-FBMC has similar BER results as conventional FBMC and CP-OFDM with SRRC $K$=8.

For SRRC *K*=4 or using other filter types, consistent with analysis, DP-FBMC has worse BER performance which is due to the higher intrinsic interference. In highly frequency selective channel vehicular B, we can see that DP-FBMC could have better performance, and this better performance is because of time multiplexing on both *Structures I* and *III*, and symbol separation on each polarization is larger than in conventional FBMC (ad CP-OFDM).

In Figure 6 plots, we compare the PSD of these three systems obtained via the periodogram technique. In Figure 6(a) we calculate these PSD results after removing the two ends of FBMC and DP-FBMC waveforms (resulting from filter tails) in order to reduce the frame lengths. We determined heuristically to truncate the first (*K*/2-1)*N* and last (*K*/2-1)*N* samples of each frame on both conventional FBMC and DP-FBMC waveforms to shorten the symbol tails due to filtering. Note that according to [31] the maximum truncation size of first and last frame samples is (*K*/2-0.25)*N* which will result to inter-frame interference in frequency selective channels, but we chose smaller truncation size to preserve the inter-frame interference as well. Based on our simulations choosing this truncation size also provide the flexibility of controlling the out of band level of PSD with the suggested SRRC filter and larger *K*. In Figure 6(a) we also show the PSD of CP-OFDM with and without windowing for comparison. In CP-OFDM windowing is used to reduce the out of band power. For the windowed CP-OFDM we used the weighted overlap and add (WOLA) based windowing technique [40] using raised-cosine window as a widely used window and roll-off factor 0.05.

According to the results and as expected, lengthening the filter (increasing *K*) using SRRC yields smaller out of band power. In Figure 6(b) we also plotted the spectra of Figure 6(a) around the band edge. As can be seen using SRRC filters also yields more compact power spectral densities. Thus after truncation, the suggested SRRC filters have more compact PSD comparing to other filters.

In Figure 7 we show BER versus carrier frequency and timing offsets (CFO, CTO) at the receiver. We compared the results with some results in the literature [41], [42] and found our results consistent for FBMC and CP-OFDM. Note that here the BER is simulated in an AWGN channel with 16-QAM modulation and $E_b/N_0$ = 12 dB with 512 subcarriers and the frame structure has 16 symbols per frame.

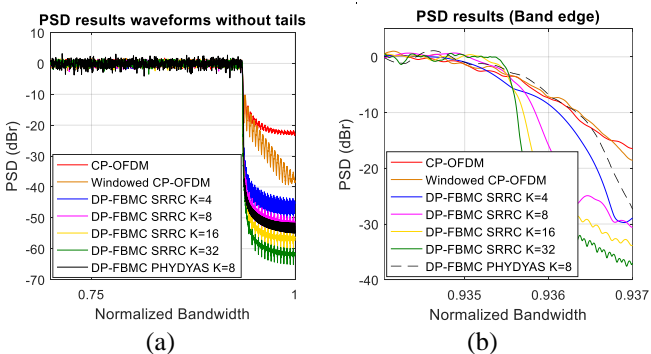

Figure 6. PSD vs. normalized bandwidth; (a) waveforms without tails, (b) around the band edge view

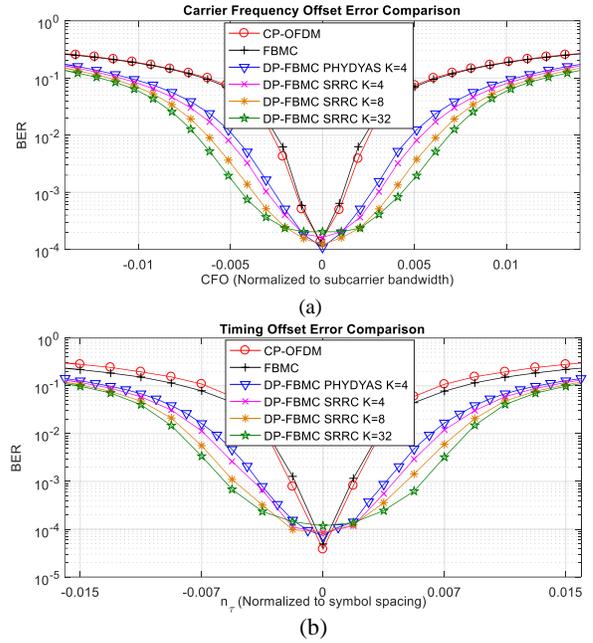

Figure 7. AWGN channel, $E_b/N_0$=12 dB, 16-QAM, 512 subcarriers, and *B*=5 MHz: (a) *BER* vs. CFO, (b) *BER* vs. CTO

The CFO values are normalized to the subcarrier bandwidth and timing offsets are normalized to DP-FBMC symbol spacing. We chose a channel bandwidth *B*=5 MHz. These results illustrate the better performance of DP-FBMC in different frequency and timing offsets. We also note that longer overlapping factors in DP-FBMC yield better BER performance versus CFO and CTO.

As more simulation results to show the effect of imperfect XPD on DP-FBMC performance we consider two scenarios. In the first scenario we assume no XP interference due to imperfect antennas or rich scattering channel environments, but instead only assume an angular mismatch between the two (*linear*) polarizations. This could be represented as the wireless communication in strong LOS channels such as AG or satellite communication. Therefore at each *θ* degree angular mismatch the received electromagnetic wave amplitudes are scaled by factors of cos(*θ*) and sin(*θ*) multiplying the desired (co-) and undesired (cross-) polarization components, respectively. Figure 8 shows the BER vs. $E_b/N_0$ results for different modulation orders in an AWGN channel (identical results for DP-FBMC *Structures I* and *III*). Using low modulation order such as QPSK, DP-FBMC has acceptable performance even at polarization angular mismatches up to 30° (approximately 1 dB loss in SNR), and this happens thanks to the π/2 phase shifts ($\theta_{n,m}$) between symbols according to (15). The theoretical results for QPSK modulation are also shown in Figure 8(a). In this case the signal to interference plus noise ratio (SINR) equals $\text{SNR} - 10\log(1 + tan^2(\theta))\ dB$ where the subtracted term is the cross-polarization interference caused by the *θ*° angular mismatch. The tolerance of the DP-FBMC system decreases for higher order modulations (results not shown here).

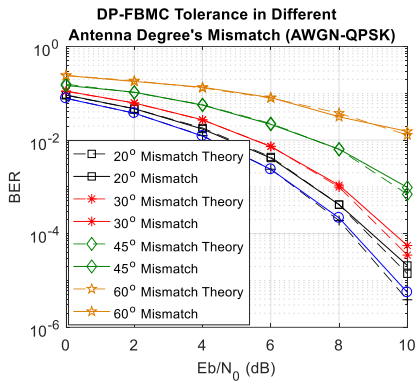

Figure 8. BER vs. $E_b/N_0$ in different angular mismatch, AWGN channel, QPSK modulation

In order to mitigate the interference from polarization mismatch we can use polarization interference cancellation techniques at the receiver. Naturally this improves performance at the expense of complexity.

In the second scenario we simulate the BER performance for XPD values from 1 to 20 dB using actual pilot-based LS channel estimation for 16-QAM. For this case we assume cross-polarization due to the channel itself and we assume no XP due to angular mismatch or imperfect antenna design. In Figure 9, simulation results for BER vs. XPD are shown for 16-QAM and two SNR values 10, and 13 dB. Here the multipath channel we used is the pedestrian channel A with bandwidth 10 MHz, and $N$=512 subcarriers. Channel equalization is based on PCK for co-polarization symbols. Other physical layer parameters are identical to those used in Figure 5. Here for the SRRC prototype filter we chose $K$=8. We also include the results assuming perfect XPD knowledge and cross-coupling interference cancelation at the receiver for comparison. As anticipated, smaller cross polarization discrimination degrades the performance, although practical XPD values of greater than 10 dB for pedestrian A channel yield performance near the ideal-XPD case. In order to enhance the performance of DP-FBMC in weak XPD conditions, as future work we could investigate a method to estimate and remove the cross-polarization interference from received signals.

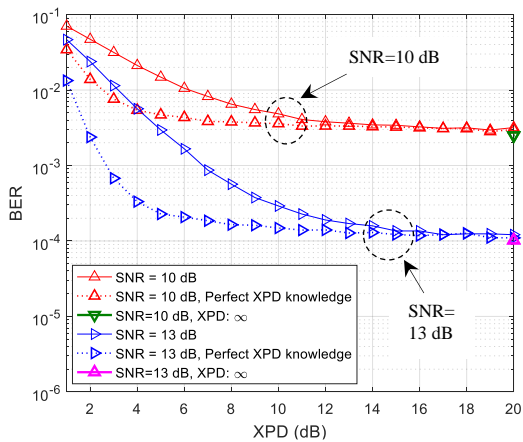

Figure 9. BER vs. XPD for 16-QAM modulation order in *ITU pedestrian channels A*, $E_b/N_0$=10, 13 dB, SRRC $K$=8, and PCK channel equalization on co-polarization symbols.

In [43] we analyzed the peak-to-average power ratio (PAPR) of DP-FBMC. According to our analysis, DP-FBMC *Structure I* has larger PAPR due to its TDM nature and temporal gaps between symbols on each polarization (other structures have similar PAPR results as conventional FBMC). As a solution, in [43] we show that using the suggested SRRC filters and larger overlapping factors in *Structure I* can also yield similar PAPR to that of CP-OFDM and conventional FBMC systems.

## V. CONCLUSION

In this paper we proposed a new FBMC system based on a dual polarization multiplexing technique. We showed that using specific time, frequency, and polarization multiplexing structures we can significantly suppress the intrinsic imaginary interference in FBMC systems. In good XPD conditions DP-FBMC provides better reliability and performance than conventional FBMC and CD-OFDM, particularly for more dispersive channels. DP-FBMC suffers in very small XPD conditions, therefore in future work we could investigate data based XPD estimation and cancellation techniques.